# The remote detectability of Earth's biosphere through time and the importance of UV capability for characterizing habitable exoplanets

**Lead author:** Christopher T. Reinhard [Georgia Institute of Technology, chris.reinhard@eas.gatech.edu]

**Co-authors:** Edward W. Schwieterman [University of California, Riverside], Stephanie L. Olson [University of Chicago], Noah J. Planavsky [Yale University], Giada N. Arney [NASA Goddard], Kazumi Ozaki [Toho University], Sanjoy Som [NASA Ames], Tyler D. Robinson [Northern Arizona University], Shawn D. Domagal-Goldman [NASA Goddard], Doug Lisman [NASA JPL], Bertrand Mennesson [NASA JPL], Victoria S. Meadows [University of Washington], Timothy W. Lyons [University of California, Riverside]

**Co-signers:** Hilairy E. Hartnett [Arizona State University], Alexander A. Pavlov [NASA Goddard], Stephen R. Kane [University of California, Riverside], Thomas Fauchez [NASA Goddard], Andrew P. Lincowski [University of Washington], Mercedes López-Morales [Harvard-Smithsonian Center for Astrophysics], Michael McElwain [NASA Goddard], Daria Pidhorodetska [NASA Goddard], Sarah Rugheimer [Oxford University], Britney Schmidt [Georgia Institute of Technology], Chuanfei Dong [Princeton University], Jade H. Checlair [University of Chicago], Peter Gao [University of California Berkeley], Jinyoung Serena Kim [University of Arizona], Ana I Gómez de Castro [Universidad Complutense de Madrid], Eric T. Wolf [University of Colorado], Wade G. Henning [University of Maryland/NASA Goddard], Wesley D. Swingley [Northern Illinois University], Nicolas Cowan [McGill University], Jacob Lustig-Yaeger [University of Washington], William Danchi [NASA Goddard], Linda Sohl [Columbia University], Alexander Pavlov [NASA Goddard], Kostas Tsigaridis [Columbia University and NASA GISS], Douglas A. Caldwell [SETI], Lars A. Buchhave [DTU Space]

**Thematic Area:** Planetary Systems

## Abstract

Thousands of planets beyond our solar system have been discovered to date, dozens of which are rocky in composition and are orbiting within the circumstellar habitable zone of their host star. The next frontier in life detection beyond our solar system will be detailed characterization of the atmospheres of potentially habitable worlds, resulting in a pressing need to develop a comprehensive understanding of the factors controlling the emergence and maintenance of atmospheric biosignatures. Understanding Earth system evolution is central to this pursuit, and a refined understanding of Earth's evolution can provide substantive insight into observational and interpretive frameworks in exoplanet science. Using this framework, we argue here that UV observations can help to effectively mitigate 'false positive' scenarios for oxygen-based biosignatures, while 'false negative' scenarios potentially represent a significant problem for biosignature surveys lacking UV capability. Moving forward, we suggest that well-resolved UV observations will be critical for near-term volume-limited surveys of habitable planets orbiting nearby Sun-like stars, and will provide the potential for biosignature detection across the most diverse spectrum of reducing, weakly oxygenated, and oxic habitable terrestrial planets.

# 1. Introduction and Relevance

We strongly endorse the findings and recommendations published in The National Academy of Sciences reports on Exoplanet Science Strategy (*ESS*) and Astrobiology Strategy for the Search for Life in the Universe (*AbS*) [*National Academy of Sciences*, 2018a,b], which represent topical and consensus statements of key scientific goals in the search for life in the cosmos [*Plavchan et al.*, 2019]. Specifically, the *ESS* stresses the importance of holistic observational approaches toward exoplanet characterization and the challenges confronting the development of theoretical frameworks for exoplanet evolution, while the *AbS* emphasizes the dynamic coevolution between life and its environment and the development of comprehensive and integrative frameworks for interpreting putative remotely detectable biosignatures. These goals are synergistic with efforts to explore the evolution of the Earth system, and share deep roots in their overarching drive to understand the stellar, biological, and planetary factors controlling habitability, atmospheric composition, and climate evolution on living worlds.

Planetary atmospheres are dynamic integrators of multiple interacting systems – including the planetary interior, surface oceans, and the evolving stellar environment – all of which can evolve dramatically across a range of timescales. The large-scale results of this complexity have been prominently featured throughout the long evolution of the only currently known inhabited planet – Earth. Indeed, the Earth system has undergone dramatic changes throughout the last ~4.5 billion years, with major transitions occurring in tectonic mode, climate dynamics, ocean chemistry, biospheric complexity, and the potential detectability of atmospheric biosignatures [*Catling and Claire*, 2005; *Korenaga*, 2006; *David and Alm*, 2011; *Lyons et al.*, 2014; *Planavsky et al.*, 2015; *Reinhard et al.*, 2017]. This complexity represents an opportunity – just as the modern Earth serves as a critical template for developing approaches toward exoplanet characterization [*Robinson et al.*, 2011; *Jiang et al.*, 2018], different periods of Earth's evolutionary history provide glimpses of largely alien worlds, some of which may be analogs for planetary states that are far more common than that of the modern Earth.

Although the James Webb Space Telescope (JWST) will allow for the characterization of exoplanets in unprecedented detail using phase curves and transit spectroscopy, the most immediate opportunity to identify life beyond our solar system will come from large ground-based observatories and future space-based direct imaging telescopes [*Mennesson et al.*, 2016; *Bolcar et al.*, 2018]. Because the science and technology definition teams (STDTs) of these mission concepts must establish instrument capabilities many years before mission launch, there is strong impetus to identify the base-level capabilities required to accomplish the scientific goals of a potential campaign in advance.

Here, we focus on the most well-developed atmospheric biosignatures for habitable terrestrial planets – molecular oxygen ($O_2$), its photochemical byproduct ozone ($O_3$), and methane ($CH_4$) – while noting that the further exploration of 'alternative' gaseous biosignatures [*Seager et al.*, 2016], surface/pigment signatures, time-variable features, and high-level interpretive frameworks remain important [*Catling et al.*, 2018; *Fujii et al.*, 2018; *Meadows et al.*, 2018; *Schwieterman et al.*, 2018]. We argue that for most of Earth's history the leveraging of well-resolved UV observations (in addition to those at optical/NIR wavelengths) would have been a critical tool for confidently detecting a surface biosphere, and that by extension UV capability should be a key consideration in the ongoing development of future direct imaging missions.

## 2. Atmospheric $O_2$ and $O_3$ abundances on Earth through time

Molecular oxygen ($O_2$) currently makes up roughly 20% of Earth's atmospheric mass, making it the second-most abundant constituent of the atmosphere. It is also spectrally active, with strong absorption features at the Fraunhofer A and B bands (0.76 and 0.69 μm, respectively) and at 1.27 μm. Earth's stratosphere contains a commensurately striking abundance of ozone ($O_3$), produced by the photolytic production and recombination of O atoms via the Chapman reactions, which also has a number of significant spectral features at ~0.35-0.20 μm (the Hartley-Huggins bands), 0.5 and 0.7 μm (the Chappuis bands), and at 9.7 μm. On the modern Earth, these atmospheric abundances are clearly sustained by biological activity, are well-mixed throughout the atmosphere and stable on geologic timescales, and are spectrally tractable through observation techniques such as transmission spectroscopy and direct imaging. Earth's atmospheric abundances of $O_2$ and $O_3$ thus satisfy the three principal criteria of a robust atmospheric biosignature – reliability, survivability, and detectability [*Meadows et al.*, 2018] – and provide strong evidence of a surface biosphere.

However, atmospheric abundances of $O_2$ and $O_3$ have changed dramatically over time as tied to a complex series of feedbacks linking Earth's tectonic evolution, ocean chemistry, nutrient abundance, and innovations in the biosphere (Fig. 1). For most of the first ~2 billion years of Earth's history atmospheric $O_2$ and $O_3$ were extremely low, and remained as low as ~$10^{-7}$ times the present atmospheric level (PAL) for perhaps as much as 700 million years after the inception of biological $O_2$ production [*Kasting et al.*, 1979; *Pavlov and Kasting*, 2002; *Claire et al.*, 2006; *Zahnle et al.*, 2006]. This Hadean/Archean Earth system state has thus been invoked as an important analog for reducing Earth-like planets [*Arney et al.*, 2016], particularly those on which oxygenic photosynthesis has not evolved [*Ozaki et al.*, 2018].

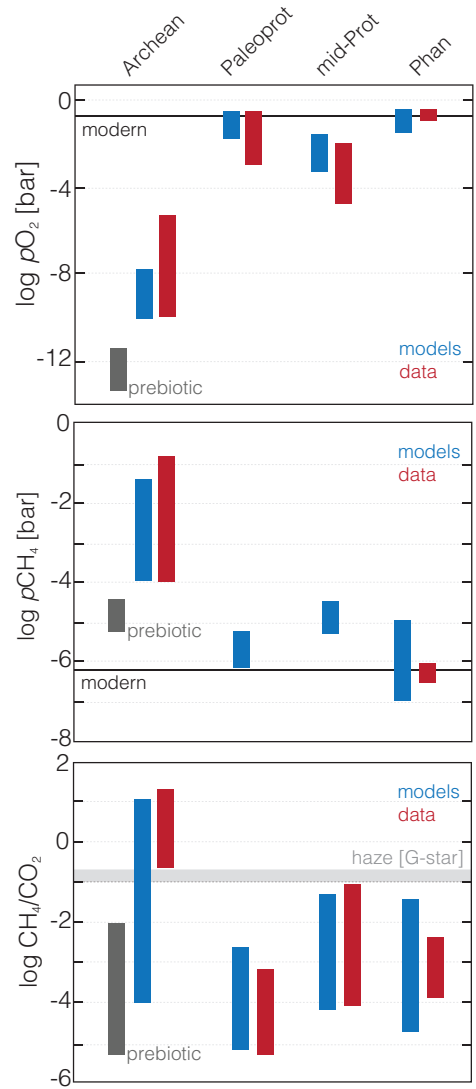

**Figure 1.** Summary of theoretical and empirical constraints on the abundances of $O_2$ (A) and $CH_4$ (B), along with the potential for atmospheric haze production (C) during four major periods of Earth's history [Hadean/Archean (~4.5-2.5 billion years ago, Ga), Paleoproterozoic (~2.5-1.8 Ga), mid-Proterozoic (~1.8-0.8 Ga), and Phanerozoic (~0.5-0 Ga)]. Blue bars show reconstructions from theoretical models, while red bars show inferences based on existing empirical data. Also shown for the Hadean/Archean are model-based estimates of prebiotic values (grey bars). The ranges are meant to be inclusive, and some of the variability in a given time period should be construed to arise from time-dependent variability rather than uncertainty. After *Robinson and Reinhard* [In review].

Isotopic evidence from Earth's rock record marks a significant and possibly rapid change in atmospheric oxidation state at ~2.3 Ga [*Farquhar et al.*, 2000; *Farquhar et al.*, 2001; *Zahnle et al.*, 2006; *Luo et al.*, 2016], but a range of geochemical constraints suggest extended periods of very low atmospheric $O_2$ and $O_3$ even after this initial rise [*Lyons et al.*, 2014; *Planavsky et al.*, 2014; *Cole et al.*, 2016; *Tang et al.*, 2016; *Bellefroid et al.*, 2018; *Planavsky et al.*, 2018]. Indeed, it is possible that atmospheric $O_2$ has been well below ~1% of the present atmospheric level for as much as 90% of Earth's evolutionary history, with roughly 'modern' atmospheric $O_2$ and $O_3$ abundances arising only in the last ~500 million years (Fig. 1).

## 3. Atmospheric $CH_4$ and organic haze abundances on Earth through time

To first order, the abundance of methane ($CH_4$) in Earth's atmosphere has inversely mirrored that of $O_2/O_3$ (Fig. 1). It is thought that during most of the Hadean/Archean atmospheric $CH_4$ abundances were ~$10^3$ times those of the modern atmosphere [*Pavlov et al.*, 2000; *Catling et al.*, 2001; *Claire et al.*, 2006; *Zahnle et al.*, 2006; *Haqq-Misra et al.*, 2008], even hundreds of millions of years after the initial evolution of oxygenic photosynthesis [*Planavsky et al.*, 2014]. These elevated atmospheric $CH_4$ abundances may have facilitated the formation of photochemical hydrocarbon 'hazes' in the Hadean/Archean atmosphere during certain periods [*Zahnle*, 1986; *Pavlov et al.*, 2001; *Domagal-Goldman et al.*, 2008; *Arney et al.*, 2016], representing a candidate biosignature that produces a very strong absorption feature at UV-blue wavelengths and responds to the magnitude of surface $CH_4$ fluxes, the background atmospheric $CH_4/CO_2$ ratio, and the production of biogenic sulfur gases [*Arney et al.*, 2018].

Because there is no direct geologic proxy for atmospheric $CH_4$ its abundance over time is extremely difficult to constrain directly beyond the ice core record [e.g., the last ~1 million years; *Higgins et al.*, 2015]. Indirect isotopic indicators of haze-induced atmospheric opacity are an exciting development [*Domagal-Goldman et al.*, 2008; *Zerkle et al.*, 2012; *Izon et al.*, 2017], but will only be sensitive when atmospheric $CH_4$ abundance is relatively high and are not currently quantitatively precise. However, existing constraints from biogeochemical models [*Daines and Lenton*, 2016; *Olson et al.*, 2016] suggest that for most of the last ~2.5 billion years atmospheric $CH_4$ abundance has been at most around an order of magnitude above that of the modern atmospheric abundance of ~$10^{-6}$ bar (Fig. 1).

## 4. False positives and false negatives

Consideration of the evolution of atmospheric $O_2/O_3$ and $CH_4$ on Earth adds to a growing literature attempting to delineate and formulate strategies to diagnose both 'false positives' – putative signatures of life that can be produced by abiotic processes – and 'false negatives' – the overprinting of biological signatures through counteracting metabolic effects or abiotic processes. With respect to potential false positives, several scenarios have been explored that may result in the abiotic accumulation of $O_2$ and $O_3$ in planetary atmospheres, most notably $O_2$ accumulation during extended pre-main sequence phase water loss around M-stars [*Luger and Barnes*, 2015] and slow hydrogen escape from thin atmospheres lacking noncondensible gases [*Wordsworth and Pierrehumbert*, 2014] (previous scenarios invoking $CO_2$ photolysis [*Tian et al.*, 2014; *Harman et al.*, 2015] have recently been shown to result from a lack of lightning-catalyzed recombination of CO and O [*Harman et al.*, 2018]). Additionally, geologic sources of $CH_4$ to the atmospheres of terrestrial planets are potentially non-trivial [*Etiope et al.*, 2011; *Etiope and Sherwood Lollar*, 2013], which could lead to buildup of abiotic $CH_4$ in reducing atmospheres. This is particularly problematic for habitable planets around M-stars, which are characterized by lower atmospheric abundances of OH and can thus support much higher atmospheric abundances of $CH_4$ at a given surface flux than Sun-like stars [*Segura et al.*, 2005; *Meadows et al.*, 2018].

Earth system evolution provides a complementary view by highlighting a number of 'false negatives' for remote life detection [*Reinhard et al.*, 2017]. For instance, it is possible that Earth's Archean atmosphere remained predominantly reducing for at least 700 million years following the evolution of oxygenic photosynthesis [*Planavsky et al.*, 2014; *Luo et al.*, 2016], with atmospheric $O_2$ and $O_3$ abundances many orders of magnitude below those that would be remotely detectable. For much of the subsequent 2 billion years atmospheric $O_2$ may have been on the order of $\sim 10^{-3}$ - $10^{-2}$ PAL, rendering the detectability of oxygen-based biosignatures possible but potentially difficult (see below). In particular, current reconstructions suggest that direct measurement of $O_2$ would have been achievable for only the last ~10% of Earth's history (Fig. 1). In addition, the detection of biogenic $CH_4$ in Earth's atmosphere, either directly or through the observation of organic-rich hazes, would have been challenging for most of the last ~2.5 billion years despite a robust and metabolically diverse biological methane cycle operating across Earth's surface since at least ~3.5 Ga [*Ueno et al.*, 2006]. In other words, much of the first half of Earth's history represents a possible false negative for oxygenic photosynthesis, while much of the latter half represents a possible false negative for biological methane cycling.

## 5. Using UV observations to mitigate false positives and false negatives

We argue here that well-resolved spectroscopic characterization at UV wavelengths is an important tool in efforts to characterize habitable terrestrial exoplanets. Such observations can potentially mitigate against both 'false positive' and 'false negative' scenarios, providing more comprehensive coverage across the range of atmospheric oxygenation states displayed throughout Earth's history. With regard to false negatives, the near-UV Hartley-Huggins feature centered at ~0.25 µm is extremely sensitive to very low levels of atmospheric $O_3$, essentially saturating when peak atmospheric $O_3$ abundance reaches ~1 ppmv. As a result, the Hartley-Huggins feature of $O_3$ would in principle allow for the detection of biogenic $O_2$ in Earth's atmosphere perhaps continuously over the last ~2.5 Ga, despite background $O_2$ levels would have rendered direct observation of molecular $O_2$ extremely challenging for all but the last ~500 million years (Fig. 1, 2). In addition, hydrocarbon haze is readily detectable via a UV-blue absorption feature, with simulations of direct imaging spectra suggesting that this feature can be confidently delineated [*Arney et al.*, 2017]. If indeed hydrocarbon haze was a common feature of Earth's Hadean/Archean atmosphere, the implication is that observation at UV wavelengths could have provided evidence of a surface biosphere for most of the last ~4.5 billion years on Earth.

Observations at UV wavelengths could also be useful in the diagnosis of oxygen-based false positives. First, the absence of particular UV/optical/NIR spectral features, such as those indicating elevated abundances of CO or $O_4$, can help rule out most false positive scenarios for M-type stars [*Schwieterman et al.*, 2016]. Second, near-UV observations can potentially help to estimate atmospheric mass via Rayleigh scattering characteristics, which would provide an important constraint on the possibility of abiotic $O_2/O_3$ buildup arising from hydrogen escape under low atmospheric pressure [*Wordsworth and Pierrehumbert*, 2014].

Lastly, direct constraints on the UV spectrum of a stellar host [e.g., *France et al.*, 2016] represent a critical boundary condition in attempts to invert atmospheric composition for surface fluxes using photochemical models. As a result, multi-wavelength star/planet characterization including UV observations together with thoughtful target selection provides strong potential to diagnose and mitigate the impact of oxygen-based false positives.

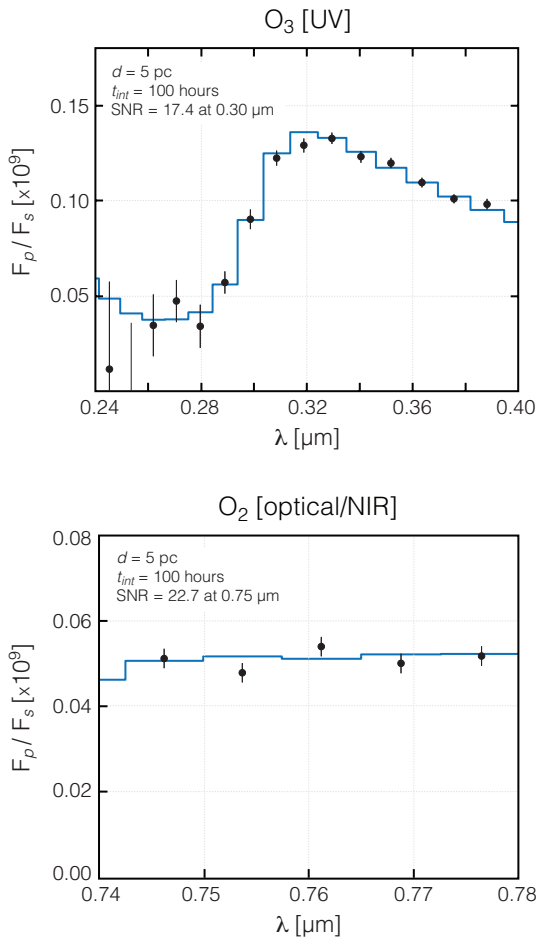

**Figure 2.** Oxygen ($O_2$) abundances that are undetectable in the optical/NIR lead to robust detection of $O_3$ in the UV. We use a radiative transfer model and a coronagraph instrument simulator [*Robinson et al.*, 2016] to model the detectability of $O_3$ in the UV (left) and $O_2$ in the optical/NIR (right) on an Earth-like planet with a weakly oxygenated atmosphere ($pO_2$ = 0.1% of the present atmospheric level, PAL) at a distance of 5 parsecs with a 10-m LUVOIR-class telescope. Planet/star contrast ($F_p/F_s$) is shown as a function of wavelength ($\lambda$). The error bars include sources of astrophysical and instrument noise including photon noise, dark current, read noise, background noise, and exozodiacal light.

In sum, although habitable terrestrial planets may undergo a wide range of evolutionary trajectories, Earth system evolution strongly suggests that UV observations provide the potential for both flexible and comprehensive mitigation against biosignature false negatives and important strengths in the diagnosis of false positives. Observations at UV wavelengths also provide the potential to increase biosignature yield via more favorable inner working angle (IWA) requirements than those of near-IR or optical observations for a wide range of architectures [*Stark et al.*, 2014; *Seager et al.*, 2015; *Robinson et al.*, 2016]. As a result, we suggest that future work should integrate simulated planetary atmospheric composition and spectral characteristics with realistic instrument performance for UV-optical-NIR-capable telescope configurations, and that this work should be coupled in real time with emerging constraints on the evolution of Earth's atmospheric composition.

## 6. Moving forward

We reiterate our endorsement of the findings and recommendations published in the National Academy reports on Exoplanet Science Strategy and Astrobiology Strategy for the Search for Life in the Universe. We extend and complement the conclusions therein by showing that UV sensitivity is critical for addressing possible false positives and false negatives in the optical/NIR reflected light spectra of habitable zone terrestrial planets. The thermal-IR spectrum of Earth also provides access to a number of gases essential for characterizing planetary climate and the possible presence of a biosphere including $CO_2$, $CH_4$, $O_3$, and $N_2O$ [*Léger et al.*, 2015; *Defrére et al.*, 2018]. In particular, the 7-8 µm $CH_4$ band produces a notable spectral impact at even modern levels and the 9.65 µm ozone band is potentially spectrally discernable at low $O_2$ levels [*Segura et al.*, 2003]. However, direct-imaging in the thermal-IR from space is not likely in the next two decades and from the ground this technique is likely limited to a handful of planets orbiting nearby M dwarf stars [*Fujii et al.*, 2018]. Transmission spectroscopy with cooled mid-IR capable space-based observatories can probe these bands as well, but only for transiting planets orbiting M and late K dwarfs, which will be on average located at greater distances than non-transiting HZ planets. Therefore, using Earth system evolution as a guide we suggest that a near-future volume-limited survey of habitable zone planets orbiting nearby Sun-like stars is best accomplished with a combined UV-optical-NIR sensitive surveyor.